\documentclass[a4paper,11pt]{article}
\pdfoutput=1
\usepackage{jheppub}
\usepackage{amsmath,amssymb,euscript}
\usepackage{slashed}
\usepackage{xspace}
\usepackage{color}
\usepackage{accents}
\usepackage{hyperref}
\usepackage{epsfig}
\usepackage{xcolor}
\usepackage{xspace}
\usepackage{verbatim}
\usepackage{multirow}
\usepackage{booktabs,graphicx}
\usepackage{mathtools}
\usepackage{tabulary}

\usepackage{graphicx}
\usepackage{caption}
\usepackage{subcaption}
\usepackage{kantlipsum}
\usepackage{caption}

\newcolumntype{K}[1]{>{\centering\arraybackslash}p{#1}}

\definecolor{cadmiumgreen}{rgb}{0.0, 0.42, 0.24}

\emergencystretch=\maxdimen
\makeatletter
\gdef\@fpheader{Presented at DIS2022: XXIX International Workshop on Deep-Inelastic Scattering and Related Subjects, Santiago de Compostela, Spain, May 2-6 2022.}
\makeatother

\begin{document}
\title{QCD at the Future Circular $\mathrm{e^{+}e^{-}}$ Collider}

\author[]{Eduardo Ploerer$^{1,2}$ \newline on behalf of the FCC collaboration} %very hacky...

\affiliation[1]{Inter-university Institute for High Energies, Vrije Universiteit Brussel, 1050 Brussels, Belgium}

\affiliation[2]{University of Zurich, 8057 Zurich, Switzerland}

\emailAdd{eduardo.ploerer@cern.ch}

%		\begin{flushright}
%		\end{flushright}

\abstract{The Future Circular Collider in the $\mathrm{e^{+}e^{-}}$ configuration offers the opportunity to significantly improve SM measurements with dedicated runs at the Z-pole, WW threshold, ZH (240 GeV), and $\mathrm{t}\bar{\mathrm{t}}$ threshold. With a factor of approximately $10^{5}$ more statistics at the Z-pole and $10^{4}$ at the WW threshold than at LEP, the FCC-ee will enable the extraction of the strong coupling $\alpha_{s}$ at the per-mille level. Parton showering studies exploiting the pure gluon sample from $\mathrm{ZH}({\rightarrow} \mathrm{gg})$ will greatly improve our understanding of gluon radiation and fragmentation, directly impacting quark vs gluon discrimination studies. Further possibilities include precision hadronization studies and colour reconnection studies at the WW threshold. Some elements of the rich precision QCD program of the FCC-ee are outlined below.}

\maketitle
%\flushbottom

\clearpage
%---------------------------------------------------------------------------
\section{Introduction}
\label{sec:intro}

Quantum Chromodynamics (QCD) is the theory of strong interactions that arises from a $SU(3)$ colour symmetry in the Standard Model (SM). The force-carrying boson is the gluon and couples to any coloured parton (itself included), making it a rather ubiquitous theory. A precise understanding of QCD is crucial for many physics measurements. In particular, a precise determination of the strong coupling $\alpha_{s}$ will improve the precision of the calculation of virtually all particle production cross sections and decays at colliders, where it would enter either directly (e.g. $\mathrm{gg} \rightarrow \mathrm{t}\Bar{\mathrm{t}}\mathrm{H}$), or via some higher-order QCD correction to a tree-level process that is purely electroweak (e.g. $\mathrm{e}^{+}\mathrm{e}^{-} \rightarrow \mathrm{Z}(\mathrm{q}\Bar{\mathrm{q}})$). For instance, the propagated uncertainty of the $\mathrm{t}\bar{\mathrm{t}}\mathrm{H}$ cross section at the LHC due to uncertainty in the strong coupling ($\delta \alpha_{s}$) is $\pm 2.2 \%$ \cite{LHCHiggsCrossSectionWorkingGroup:2016ypw}, while the main source of parametric uncertainty in the determination of the Z boson width at the FCC-ee will be $\delta \alpha_{s}$ \cite{Freitas:2019bre}. Highly related is the computation of higher-order $\mathrm{N}^{n}\mathrm{LO}$ corrections and $\mathrm{N}^{n}\mathrm{LL}$ resummations. Such corrections arise in any perturbative QCD (pQCD) expansion, and as such are central to increasing the precision in the prediction of observables. A precise determination of Parton Distribution Functions (PDFs) is necessary to compute production cross sections at pp colliders. The present uncertainty in the $\mathrm{t\Bar{t}H}$ cross section due to PDFs is $\pm 3.9 \%$ \cite{LHCHiggsCrossSectionWorkingGroup:2016ypw}. The identification and separation of heavy quarks, light quarks, and gluons relies on a precise picture of jet substructure and parton showering, and would enable the enhancement of jet signals, particularly in pp collisions where one has to contend with a sizeable gluon jet background. Other non-pQCD phenomena include hadronization, whose precise understanding benefits any hadronic final state, and colour reconnection.

The FCC-ee \cite{FCC:2018} is typically not considered a QCD machine, but offers a rich QCD programme owing to its clean environment and huge integrated luminosities \cite{Skands:2016}. The collision of colourless partons not only fixes the initial state kinematics in 3 dimensions (as opposed to pp where they are only fixed in the transversely allowing for longitudinal boosts), but also removes the necessity of dealing with PDFs. Indeed, the lack of QCD initial state radiation (ISR), multiple parton interactions (MPIs), and pileup means that quarks and gluons hadronize in a relative “QCD vacuum”. Hence, overlap between jets is considerably reduced, resulting in much more well-defined jets than in pp collisions. Combining the clean environment with the large expected number of events at the FCC-ee of roughly $\sim 10^{11} \mathrm{e^{+}e^{-} \rightarrow  Z(q\bar{q})}$ per quark flavour at $\sqrt{s}\sim$ 91 GeV, $\sim 10^{7} \; \mathrm{e^{+}e^{-}} \rightarrow \mathrm{W^{\ast}W(q_{1}q_{2})}$ per quark flavour $\mathrm{q_{1}}$ at $\sqrt{s} \sim$ 160 GeV, and $\sim 10^{(6, 5, 5)} \; \mathrm{e^{+}e^{-} \rightarrow ZH(b\bar{b},c\bar{c},gg)}$ at $\sqrt{s} \sim$ 240 GeV provides data sets for unprecedented precision QCD studies. Other future $\mathrm{e^{+}e^{-}}$ colliders offer similarly clean environments, with designs being either circular (as the FCC-ee and CEPC \cite{CEPCStudyGroup:2018rmc}) or linear (as the ILC \cite{ILCTDR:2013xla} and CLIC \cite{CLICCDR:2012hp}). Circular colliders feature larger luminosities at lower $\sqrt{s}$, while linear colliders feature a higher center of mass energy range. 

\section{QCD coupling $\mathbf{\alpha_{s}}$}
\label{sec:alpha_s}

The extraction of the strong coupling $\alpha_{s}$ has a long history and is still a very active endeavor. In 1989 the uncertainty on $\alpha_{s}$ was $\delta\alpha_{s}/\alpha_{s} \sim$ 6 \% \cite{Altarelli:116932}, which decreased to $\delta\alpha_{s}/\alpha_{s} \sim$ 2.5 \% in 2000 \cite{ParticleDataGroup:2000nwm}, and $\delta\alpha_{s}/\alpha_{s} \sim$ 0.85 \% in 2019 \cite{ParticleDataGroup:2020ssz}. 

Nevertheless, $\alpha_{s}$ remains the least precisely known of all interaction couplings with $\delta \alpha_{QED}\sim 10^{-10} \ll \delta G_{F} \sim 10^{-7} \ll \delta G \sim 10^{-5} \ll \delta \alpha_{s} \sim 10^{-3}$. Presently, $\alpha_{s}$ is determined by comparing 7 experimental observables to their pQCD predictions. A global average is performed at $\alpha_{s}(m_{\mathrm{Z}})$. The observables that pertain to $\mathrm{e^{+}e^{-}}$ collisions are hadronic $\tau$ lepton decays, $\mathrm{e^{+}e^{-}}$ jet shapes and rates,  and hadronic Z boson decays (in the future also hadronic W boson decays and $\mathrm{e^{+}e^{-} \rightarrow t\bar{t}}$ cross sections) \cite{dEnterria:2022hzv}.

\subsection{$\mathbf{\alpha_{s}}$ from hadronic $\mathbf{\tau}$-lepton decays}

The extraction of $\alpha_{s}$ from hadronic $\tau$ lepton decays can be done by exploiting the precise LEP and B-factories $\mathrm{e^{+}e^{-}} \rightarrow \tau^{+}\tau^{-}$ data and higher-order pQCD corrections to the hadronic $\tau$ width. The ratio of the hadronic $\tau$ width and the electron $\tau$ width is
\begin{equation}
R_{\tau} = \frac{\Gamma \left(\tau^{-} \rightarrow \nu_{\tau} + \mathrm{hadrons}\right)}{\Gamma\left(\tau^{-} \rightarrow \mathrm{e}^{-}\bar{\nu}_{\mathrm{e}} \nu_{\tau}\right)} = S_{\mathrm{EW}}N_{\mathrm{C}}\left(1+\Sigma^{4}_{n=1} c_{n} \left(\frac{\alpha_{s}}{\pi}\right)^{n} + \mathcal{O}\left(\alpha_{s}^{5}\right) + \delta_{\mathrm{np}}\right)
\end{equation}
where $S_{EW}$ is the pure electroweak contribution to the ratio, $N_{\mathrm{C}}$ is the number of colours, $c_{n}$ are coefficients of the perturbative expansion, and $\delta_{\mathrm{np}}$ are power suppressed non-pQCD corrections. The ratio is preferable to the hadronic branching width itself as several systematic uncertainties cancel between the numerator and denominator. Presently, $R_{\tau , \mathrm{exp.}} = 3.4697 \pm 0.0080$ is determined experimentally to $\pm 0.23 \%$ precision. The experimental measurement can be compared to the $\mathrm{N^{3}LO}$ theoretical prediction to extract $\alpha_{s}(m_{\mathrm{Z}}) = 0.1187 \pm 0.0018$ at the $\pm 1.5 \%$ level.

Theoretically, the dominant source of uncertainty in the determination of $\alpha_{s}(m_{\mathrm{Z}})$ is the discrepancy between two different approaches for evaluating the perturbative expansion: contour-improved perturbation theory (CIPT) and fixed-order perturbation theory (FOPT). Discrepancies between FOPT and CIPT are actively being investigated \cite{Benitez-Rathgeb:2021gvw} as is evidenced in the uncertainty increase from $\pm 1.3 \%$ in 2013 to $\pm 1.5 \%$ in 2019.

Non-perturbative corrections are non-negligible in the determination of $\alpha_{s}$ from hadronic $\tau$ lepton decays. These are power suppressed as $\mathcal{O}(\Lambda^{p}/Q^{p})$ where $\Lambda$ is the QCD scale parameter and $Q$ is the momentum transfer, and as such can be sizeable for $\mathcal{O}(\Lambda_{\mathrm{QCD}}^{2}/m_{\tau}^{2}) \sim 2 \%$. Controlling these corrections requires new high-precision measurements of the hadronic tau spectral function.

At the FCC-ee the production of $10^{11}$ $\tau$ leptons at the Z-pole means that the statistical uncertainty will be negligible. Systematic and parametric uncertainties will dominate. A reduction in the spread of theoretical determinations of $R_{\tau}$ is of paramount importance to a full exploitation of the FCC-ee statistics. Central to this is a better understanding of FOPT and CIPT discrepancies. Moreover, improvements in the determination of spectral functions entering $R_{\tau}$ are necessary, which could come from Belle II or the FCC-ee itself. In this way the uncertainty on $\alpha_{s}$ could be reduced well below the current $\delta\alpha_{s}/\alpha_{s} \sim 1\%$ level.  

\subsection{$\mathbf{\alpha_{s}}$ from e$^{+}$e$^{-}$ event shapes \& jet rates}

Another way to measure $\alpha_{s}$ is through the measurement of event shapes and jet rates. The thrust and the $C$-parameter defined below were used to extract $\alpha_{s}$ at LEP \cite{Dissertori:2009ik},
\begin{align}
    \tau  = 1-T = 1-\mathrm{max}_{\hat{n}}\frac{\Sigma|\Vec{p}_{i}\cdot\hat{n}|}{\Sigma|\Vec{p}_i|} && C  = \frac{3}{2}\frac{\Sigma_{i,j}|\Vec{p}_{i}||\Vec{p}_{j}|sin^{2}\theta_{ij}}{(\Sigma_i|\Vec{p}_i|)^{2}}
\end{align}
where $\Vec{p}_{i}$ is the $i^{th}$ particle's 4 momentum and $\theta_{ij}$ is the angle between particles $i$ and $j$. In addition, the  n-jet rates $R_{n} = \frac{\sigma_{njet}}{\sigma_{tot.}}$   are sensitive to $\alpha_{s}$ and were used to extract $\alpha_{s}$. A comparison of the experimental measurements to $\mathrm{N^{2,3}LO}$+$\mathrm{N^{(2)}LL}$ predictions yields $\alpha_{s}(m_{\mathrm{Z}}) = 0.1171 \pm 0.0027$ ($\pm 2.6\%$).

The theoretical predictions of event shapes and jet rates are especially sensitive to non-pQCD. These effects are modelled via MC generators or analytically with some disagreement, leading to a considerable spread in predictions.

At the FCC-ee the tagging of ISR could be used to study the energy dependence of jet shape observables at lower $\sqrt{s}$. At the Z-pole n-jet rates of up to 7 (with $k_{\mathrm{T}} \approx 7.5$ GeV) could be studied. Moreover, runs at higher $\sqrt{s}$ could be used to study jet rates in regimes where the probability of hard gluon emission increases. Improvements in logarithmic resummation to $\mathrm{N^{2,3}LL}$ for jet rates and the understanding of hadronization for event shapes would enable the extraction of of $\alpha_{s}$ below the $\delta\alpha_{s}/\alpha_{s} < 1\%$ level at the FCC-ee.

\subsection{$\mathbf{\alpha_{s}}$ from hadronic Z boson decays}
Analogously to the case for hadronic $\tau$ lepton decays, $\alpha_{s}$ can be extracted from hadronic Z boson and W boson decays by considering the ratio of the hadronic width to the lepton width: 
\begin{equation}
R_{\mathrm{Z},\mathrm{W}}\left(Q\right) = \frac{\Gamma_{\mathrm{Z},\mathrm{W}}^{\mathrm{had.}}\left(Q\right)}{\Gamma_{\mathrm{Z},\mathrm{W}}^{\mathrm{lep.}}\left(Q\right)} = R_{\mathrm{Z},\mathrm{W}}^{\mathrm{EW}}\left(1+\Sigma^{4}_{i=1} a_{i}\left(Q\right) \left(\frac{\alpha_{s}(Q)}{\pi}\right)^{i} + \mathcal{O}\left(\alpha_{s}^{5}\right) + \delta_{\mathrm{mix}} + \delta_{\mathrm{np}}\right)
\end{equation}
where $R_{\mathrm{Z},\mathrm{W}}^{\mathrm{EW}}$ is the pure electroweak contribution to the ratio, $a_{i}$ are coefficients of the perturbative expansion, $\delta_{\mathrm{mix}}$ are mixed QCD+EW corrections, and $\delta_{\mathrm{np}}$ are power suppressed non-pQCD corrections. The strong coupling can then be extracted by performing a simultaneous fit of three Z boson pseudo-observables \cite{dEnterria:2020cpv-ins}: $R_{\mathrm{Z}}$, $\Gamma_{\mathrm{Z}}^{\mathrm{tot.}}$, $\sigma_{\mathrm{Z}}^{\mathrm{had.}}$ yielding $\alpha_{s} = 0.1203 \pm 0.0029$. A full SM fit can be performed yielding $\alpha_{s} = 0.1203 \pm 0.0029$ ($\pm 2.3 \%$), as depicted in Figure \ref{Zfig}.

At the FCC-ee the production of $10^{5}$ times more Z bosons than at LEP and exquisite systematic and parametric precision means that improvements in theoretical predictions of Z boson pseudo-observables would enable a reduction in the uncertainty of $\alpha_{s}$ by an order of magnitude. Such an experimental precision needs to be matched by reducing the theoretical uncertainty by a factor of 4 by computing missing $\alpha_{s}^{5}$, $\alpha^{3}$, $\alpha\alpha_{s}^{2}$, $\alpha^{2}\alpha_{s}$ which results $\alpha_{s}(m_{\mathrm{Z}}) = 0.12030 \pm 0.00026$ ($0.2 \%$) \cite{dEnterria:2020cpv-ins} at the FCC-ee, as depicted in Figure \ref{Zfig}.

\subsection{$\mathbf{\alpha_{s}}$ from hadronic W boson decays}
\begin{figure}[h]
  %\centering
\begin{minipage}[t]{0.48\textwidth}
\centering
\includegraphics[width=0.97\linewidth]{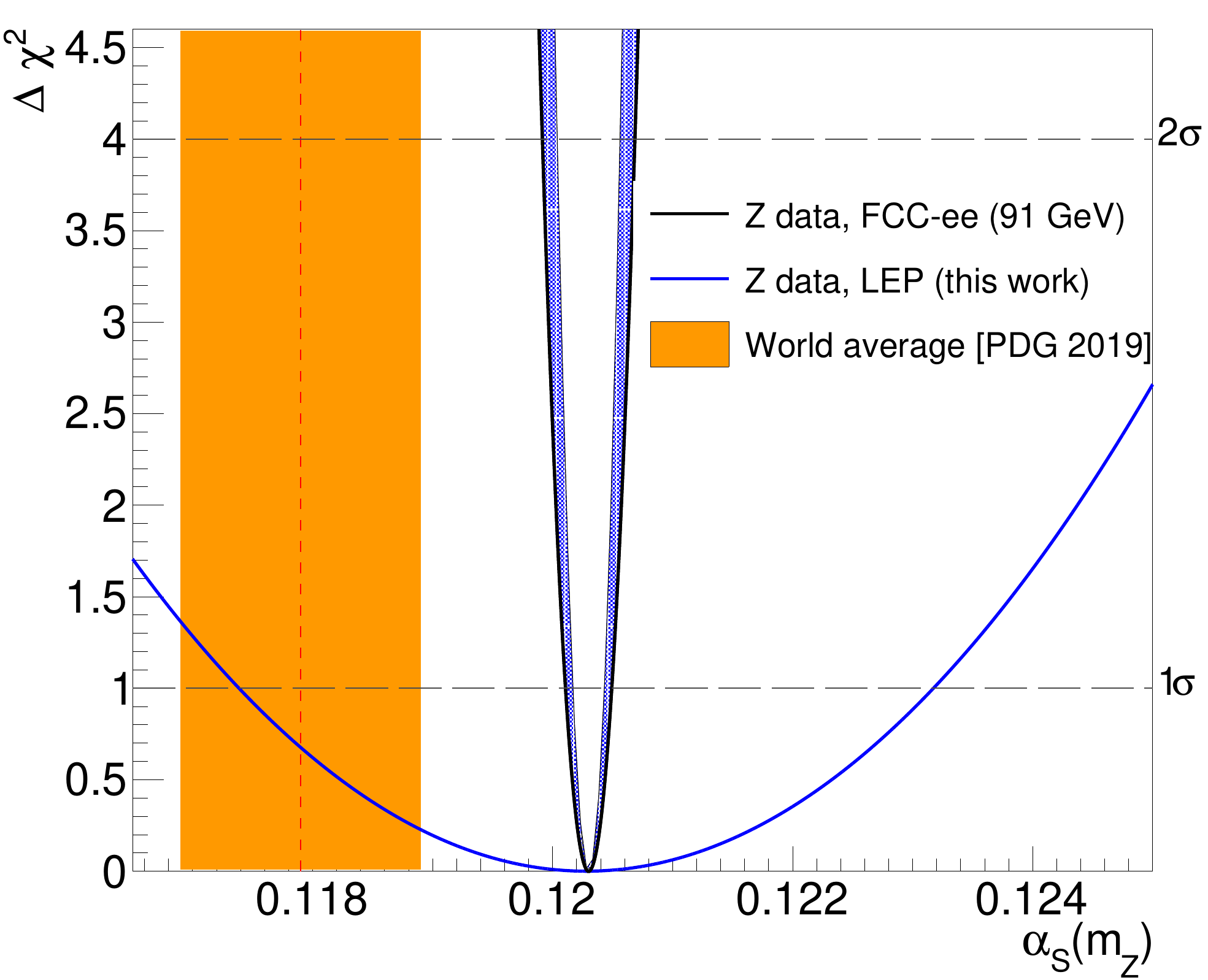}
\caption{$\Delta \chi^{2}$ fit profile of $\alpha_{s}(m_{\mathrm{Z}})$ extracted from Z boson pseudo-observables ($R_{\mathrm{Z}}$, $\Gamma_{\mathrm{Z}}^{\mathrm{tot.}}$, $\sigma_{\mathrm{Z}}^{\mathrm{had.}}$) from LEP data (blue curve), and from expected FCC-ee data (black parabolas) with central value arbitrarily chosen. The outer parabola corresponds to the sum of experimental, parametric, and theoretical uncertainties in quadrature, while the inner parabola corresponds to purely the experimental uncertainty. The world average (dashed red line) is depicted with its uncertainty (orange band). Figure taken from \cite{dEnterria:2020cpv-ins}.}
\label{Zfig}
\end{minipage}
\vspace{0pt}
\hfill
\begin{minipage}[t]{0.48\textwidth}
\centering
\includegraphics[width=0.97\linewidth]{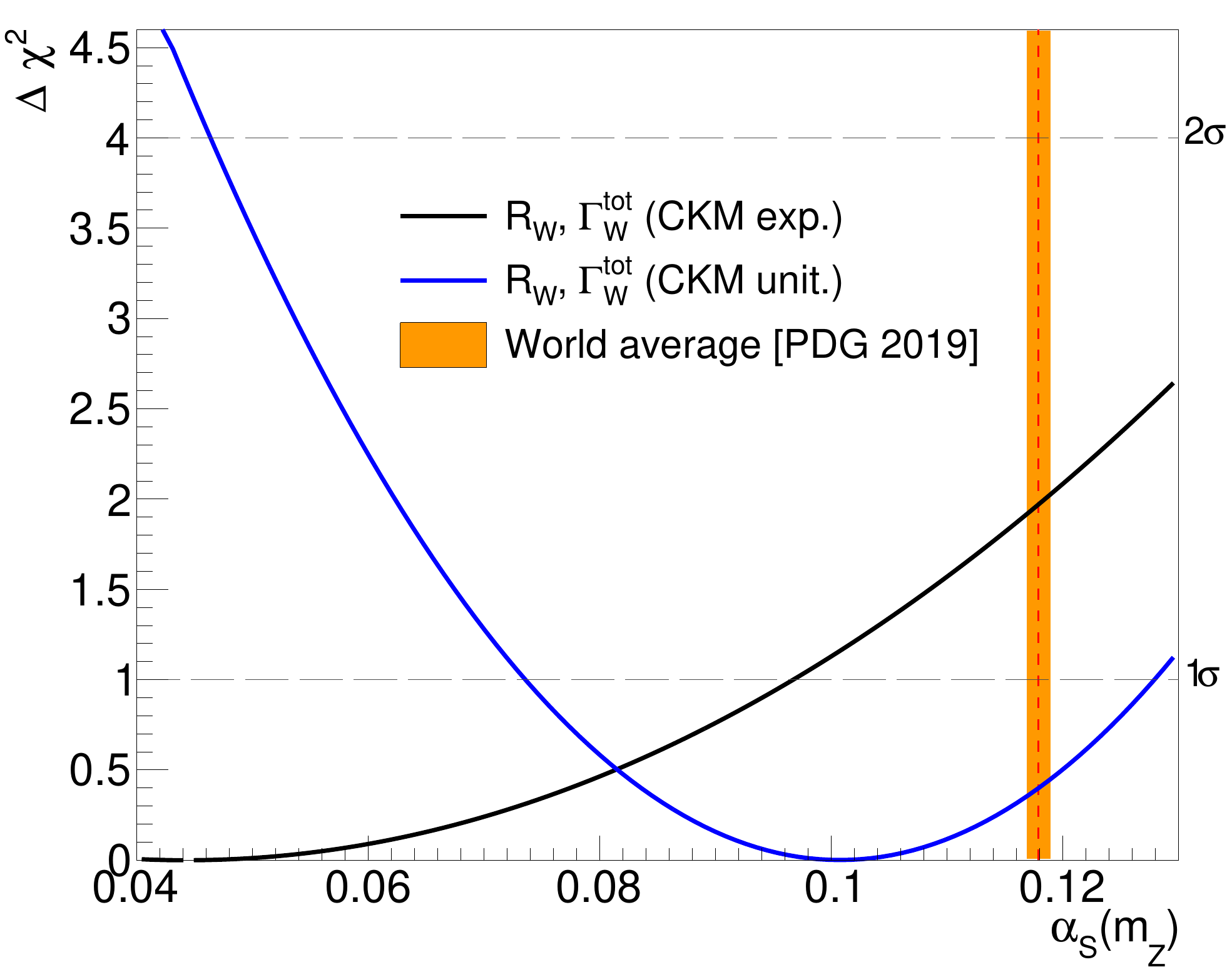}
\caption{$\Delta \chi^{2}$ fit profile of $\alpha_{s}(m_{\mathrm{Z}})$ extracted from W boson pseudo-observables ($R_{\mathrm{W}}$ and $\Gamma_{\mathrm{W}}$) from LEP data with (blue curve) and without (black curve) the assumption of CKM unitarity. The world average (dashed red line) is depicted with its uncertainty (orange band). Figure taken from \cite{dEnterria:2020cpv-ins}. \bigskip \smallskip}
\label{Wfig}
\end{minipage}
\end{figure}

The extraction of $\alpha_{s}$ from hadronic W boson decays can be done by performing a simultaneous fit of $\mathrm{N^{3}LO}$ predictions of $R_{\mathrm{W}}$ and $\Gamma_{\mathrm{W}}^{\mathrm{tot.}}$ to their measured values \cite{dEnterria:2020cpv-ins}. Without the assumption of CKM unitarity this leaves the strong coupling unconstrained at $0.04 \pm 0.05$. This is largely due to the significant experimental uncertainty of $V_{\mathrm{cs}}$ of roughly $\pm 2 \%$. Assuming CKM unitarity one arrives at $\alpha_{s} = 0.101 \pm 0.027$ ($\pm 27\%$), as depicted in Figure \ref{Wfig}. The large uncertainty can be attributed to poor experimental precision of $R_{\mathrm{W}}$ and $\Gamma_{\mathrm{W}}$ measured in $\mathrm{e^{+}e^{-}} \rightarrow \mathrm{W^{+}W^{-}}$ collisions at LEP. The corresponding theoretical uncertainty of $\sim 1.5 \%$ is much smaller than the experimental uncertainties.

\begin{figure}[h]
  %\centering
\begin{minipage}[t]{0.48\textwidth}
\centering
\includegraphics[width=0.97\linewidth]{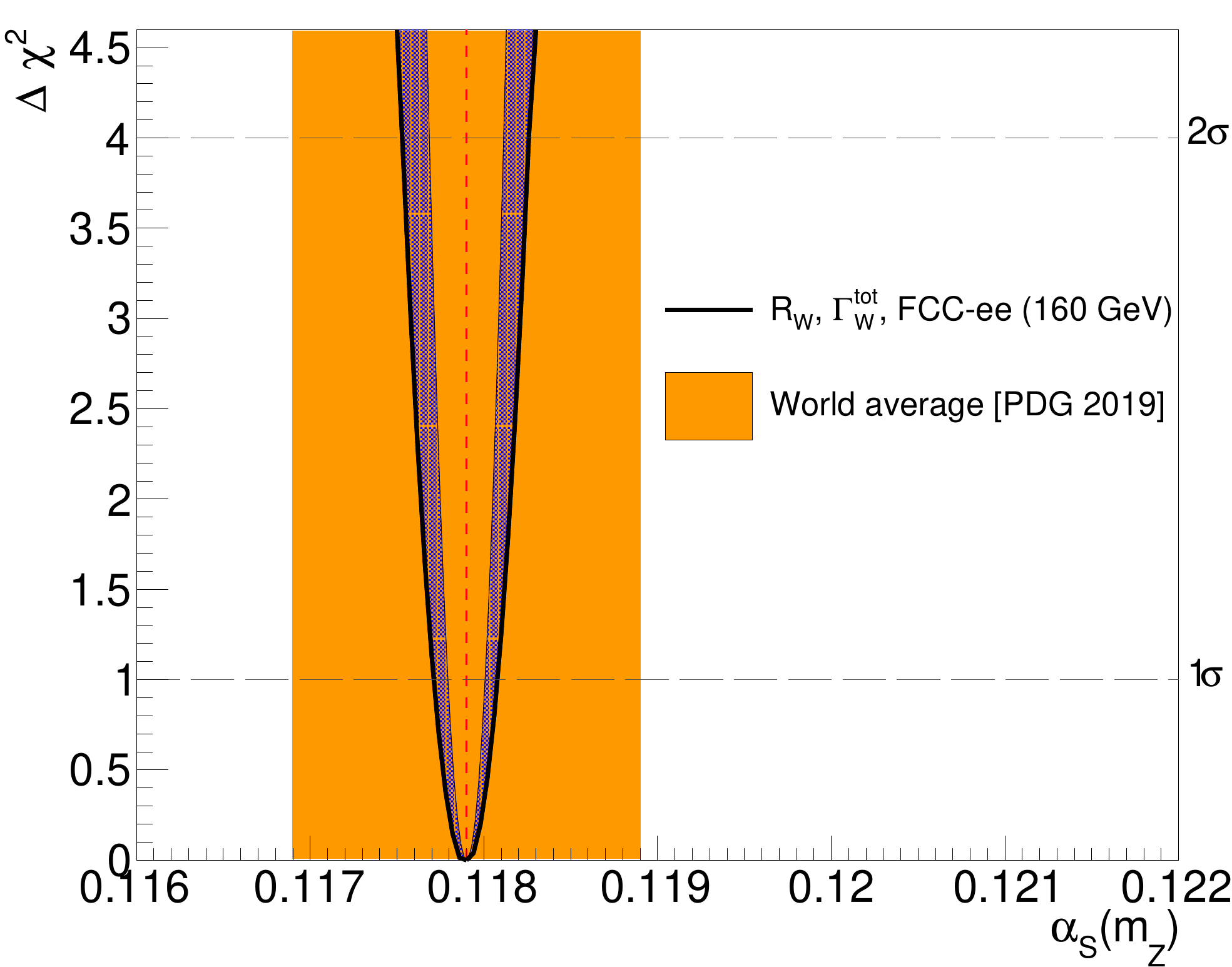}
\caption{$\Delta \chi^{2}$ fit profile of $\alpha_{s}(m_{\mathrm{Z}})$ extracted from W boson pseudo-observables ($R_{\mathrm{W}}$ and $\Gamma_{\mathrm{W}}$) from expected FCC-ee data with only experimental (inner parabola) and experimental, parametric, and theoretical uncertainties (outer parabola) with the assumption of CKM unitarity. The world average (dashed red line) is depicted with its uncertainty (orange band). Figure taken from \cite{dEnterria:2020cpv-ins}.}
\label{WFCCfig}
\end{minipage}
\vspace{0pt}
\hfill
\begin{minipage}[t]{0.48\textwidth}
\centering
\includegraphics[width=1.07\linewidth]{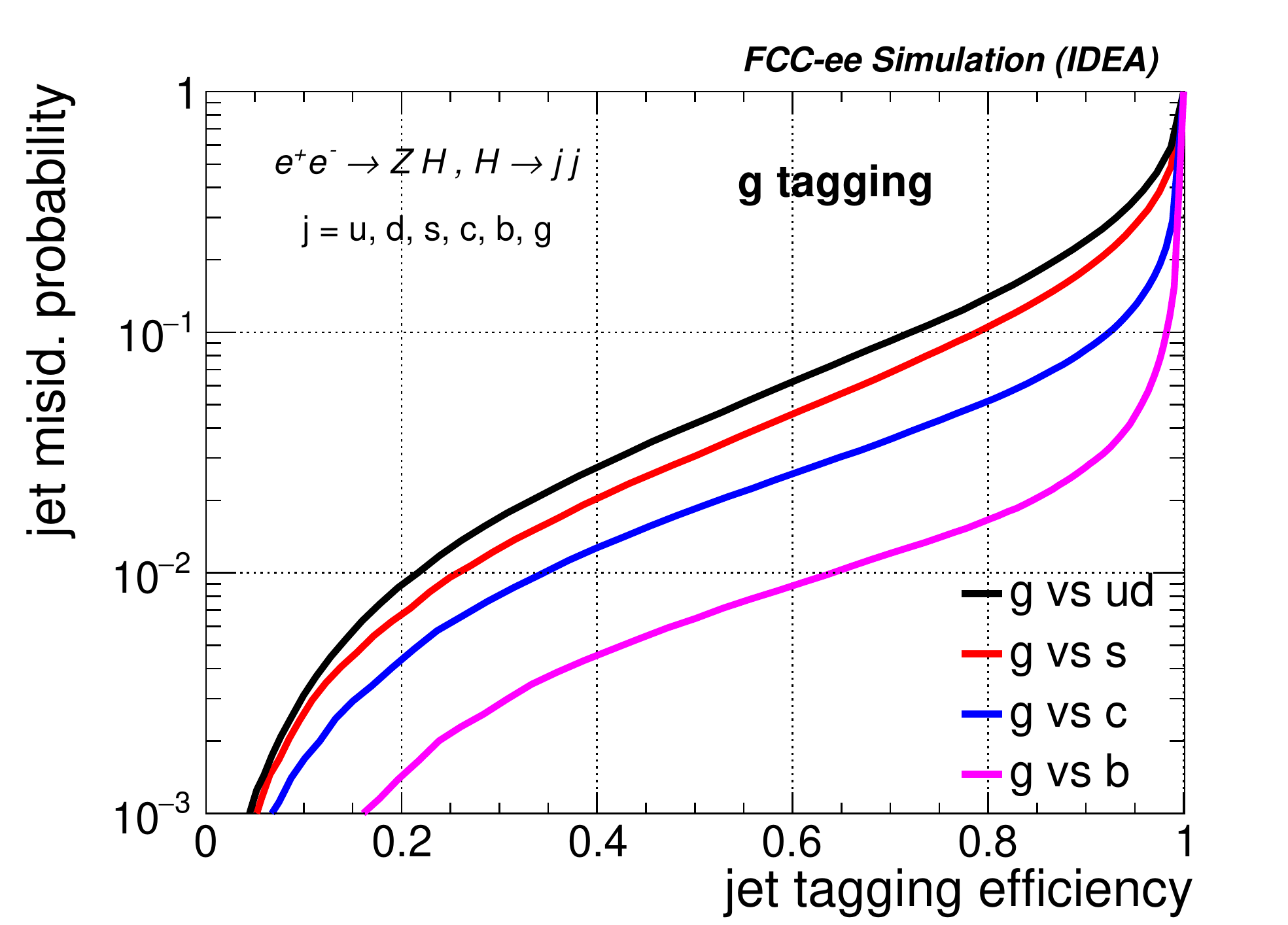}
\caption{Gluon jet classification performance of ParticleNetIdea as receiver operating characteristic (ROC) curves in FCC-ee fast simulation using the IDEA detector concept \cite{FCC:2018, Bedeschi:2021bS}. In each curve different quark flavours are considered background: u+d quarks (black), s quarks (red), c quarks (blue), and b quarks (pink).  Figure taken from \cite{Bedeschi:2022rnj}.}
\label{ParticleNetIdea}
\end{minipage}
\end{figure}

At the FCC-ee the large statistics at the $\mathrm{WW}$ threshold mean that the uncertainties of $R_{\mathrm{W}}$, $\Gamma_{\mathrm{W}}^{\mathrm{tot.}}$ will be considerably reduced. The corresponding theoretical uncertainty can be reduced by a factor of 10 by computing missing $\alpha_{s}^{5}$, $\alpha^{3}$, $\alpha\alpha_{s}^{2}$, $\alpha^{2}\alpha_{s}$ terms. The extraction of the strong coupling could then improve by two orders of magnitude to $\alpha_{s}(m_{\mathrm{Z}}) = 0.11790 \pm 0.00023$ ($0.2 \%$) \cite{dEnterria:2020cpv-ins}, as depicted in Figure \ref{WFCCfig}. 

\section{Quark-gluon tagging \& jet substructure}
\label{sec:qg_tagging_jet_subs}

Light quark-gluon discrimination is an exciting but challenging prospect in pp collisions. Distinguishing gluon initiated from light quark initiated jets could enhance light quark-rich signals such as in Vector Boson Fusion, $\mathrm{t\bar{t}H}$ production (hadronic $\mathrm{W}$ bosons), pure electroweak $\mathrm{W}$/$\mathrm{Z}$ + jets, etc. Moreover, accurate light quark-gluon discrimination would open the door to the study of BSM physics signals without leptons, b or top quarks. 

There exist several handles to experimentally separate jets initiated by light quarks and by gluons \cite{JetSubsSchwartz:2013}. The most significant discriminating variables arise from the differing colour factors for quarks and gluons: $C_{F} = 4/3$ vs $C_{A} = 3$. These govern the emission probabilities of gluons at qqg and ggg vertices, respectively. Hence, gluon jets radiate more, tending to be wider and having more constituents. 

The differing spins of quarks and gluons give rise to spin correlations in subjet locations. In particular, the production of identical quarks overlapping in spacetime (e.g. along axis of quark from hard scattering) is Fermi-Dirac suppressed, while the production of gluons overlapping in spacetime is Bose-Einstein enhanced.

The transferring of the initial parton's charge through hadronization can also be exploited via variables such at the $p_{\mathrm{T}}$-weighted jet charge. The usefulness of this variable is limited by the fractional charges of quarks. 

A multitude of machine learning approaches \cite{PARTICLENET:2020, DeepJet:2020, DLColour:2017, EFN:2019, ATLASQG:2017, CMSQG:2017} have recently exploited the above variables and found success in discriminating light quark and gluon jets. In particular, ParticleNetIdea \cite{Bedeschi:2022rnj} has been developed in the context of the FCC-ee, and was able to discriminate light quark jets from gluons with high efficiency, as depicted in Figure \ref{ParticleNetIdea}. Nevertheless it remains an open question how much one can trust such approaches. Machine learning is known to be especially sensitive to mismodelling of input variables, and discriminating variables rely on a precise understanding of non-pQCD phenomena. 

Jet substructure studies will be central to improving the modelling of parton showering and hadronization. An attractive starting point for such studies are jet angularities \cite{Larkoski:2014} defined as $\lambda^{\kappa}_{\beta} = \Sigma_{i \in \mathrm{jet}} z_{i}^{\kappa} \theta_{i}^{\beta}$, where $z_{i}$ and $\theta_{i}$ are the energy fraction and angular distance to jet axis of constituent $i$. The parameters $\kappa \geq 0$ and $\beta \geq 0$ govern the energy and angular weighting, respectively. Common examples of jet angularities include the multiplicity ($\kappa=0$, $\beta=0$), width ($\kappa=1$, $\beta=1$), mass ($\kappa=1$, $\beta=2$), $p_{\mathrm{T}}^{\mathrm{D}}$ ($\kappa=0$, $\beta=2$), Les Houches Angularity ($\kappa=1$, $\beta=0.5$). Fixing $\kappa = 1$ results in an infrared- and collinear-safe set of variables.

The Les Houches Angularity in particular offers an opportunity to study showering differences between generators. It was not directly measured at LEP, despite multiple generators being tuned to LEP data. Comparing quark and gluon radiation patterns between different generators in $\mathrm{e}^{+}\mathrm{e}^{-}\rightarrow \mathrm{Z}\rightarrow \mathrm{u}\mathrm{u}$ and $\mathrm{e}^{+}\mathrm{e}^{-}\rightarrow \mathrm{H} \rightarrow \mathrm{gg}$ shows considerable differences for gluons \cite{GSoyez:2017}. This manifests as vastly different discriminating power between quarks and gluons depending on the generator used. 

The FCC-ee would be crucial in addressing such modelling shortcomings. The 120k $\mathrm{e^{+}e^{-}} \rightarrow \mathrm{ZH}(\rightarrow \mathrm{gg})$ events could be used to study gluon radiation patterns. These could be contrasted to light quark initiated jets from $\mathrm{e^{+}e^{-}} \rightarrow \mathrm{Z}\rightarrow \mathrm{q\bar{q}}$ events. Another source of pure gluon jets is $\mathrm{e^{+}e^{-}} \rightarrow \mathrm{Z}\rightarrow \mathrm{b\bar{b}g}$ where the b jets are tagged with high efficiency. In this way jet angularity studies at the FCC-ee would lead directly to improved MC tuning, improvements in q vs g discrimination tools, and improved non-pQCD understanding as outlined below. 

\section{Non-perturbative QCD}
\label{sec:non_pQCD}

Colour reconnection is a set of phenomena impacting the final state kinematics of hadronic states in the form of shifted angular correlations and/or invariant mass shifts \cite{Christiansen:2015yca}. In the context of string hadronization, colour reconnection typically refers to a rearrangement of the hadronizing colour singlets. Such effects can be associated with soft gluon exchanges arising from a departure from the “leading colour limit” ($N_{\mathrm{C}}\rightarrow \infty$) during hadronization, where the $\frac{1}{N_{\mathrm{C}}}$ suppressed term in the Fierz identity is neglected \cite{Bierlich:2022pfr}. As such they are suppressed by $(\frac{1}{N_{\mathrm{C}}})^{2} \sim 10 \%$. The exact dynamics are poorly understood, and its modelling is an active field. 

Nevertheless, colour reconnection can lead to sizable uncertainties when the hadronization of separate colour singlets is overlapping in spacetime. This is the case for hadronic $m_{\mathrm{W}}$, $m_{\mathrm{top}}$ measurements, or anomalous gauge coupling extractions. For instance, in $\mathrm{e^{+}e^{-}} \rightarrow \mathrm{W(q_{1}\bar{q}_{2})W(q_{3}\bar{q}_{4}})$ since $\Gamma_{\mathrm{W}} \gg \Lambda_{\mathrm{QCD}}$ both W bosons will decay before they have travelled far enough to prevent their hadronization regions from overlapping. Then the hadronizing singlets may flip from ($\mathrm{q_{1}\bar{q}_{2}}$), ($\mathrm{q_{3}\bar{q}_{4}}$) to ($\mathrm{q_{1}\bar{q}_{4}}$), ($\mathrm{q_{3}\bar{q}_{2}}$) leading to the “string-drag effect” on the W mass hinted at at LEP. LEP was able to exclude the no-colour reconnection hypothesis at the $99.5\%$ CL, but could not constrain colour reconnection further \cite{LEPCR:2013}. 

At the FCC-ee colour reconnection will impact a number of multi-jet hadronic states including $\mathrm{e^{+}e^{-}}\rightarrow \mathrm{WW}$, $\mathrm{e^{+}e^{-}}\rightarrow \mathrm{t\bar{t}}$, $\mathrm{H(WW)}$, etc. In particular, at the $\mathrm{WW}$ threshold the huge ($\times10^{4}$ LEP) statistics could be exploited to measure $m_{\mathrm{W}}$ hadronically and (semi-) leptonically to contrain colour reconnection at the $1 \%$ level or below. Colour reconnection studies would especially benefit from the absence of MPIs present in pp-collisions, which can complicate hadronization considerably through “cross-talk” with partons not in the hard scattering. 

With the particle type identification of charged hadrons expected at the FCC-ee a multitude of hadronization studies would be possible. These would be centered around the study of colour string dynamics and in particular the conservation of baryon and strangeness numbers. Moreover, final-state correlations such as spin or momentum correlations arising from Bose-Einstein enhancement or Fermi-Dirac suppression would be studied. The formation of quark and gluon bound states such as quarkonia, multi-quark states, glueballs could likewise be studied. 
The apparent breakdown in the universality of hadronization observed at ALICE \cite{ALICE:2017} raises several important questions. The FCC-ee could serve as an environment to study hadronization in a “QCD vacuum”, and thus serve as the baseline for high-density QCD studies, where one would have to contend with considerably more non-trivial final-state medium effects such as a quark gluon plasma.  

\section{Conclusion}

A precise understanding of pQCD and non-pQCD physics is needed to fully exploit future collider programs. A plethora of unique QCD studies are possible at the FCC-ee. These include the per mille extraction of $\alpha_{s}$ via hadronic tau/Z/W decays, event shapes, etc. Jet substructure studies could be matched to $\mathrm{N}^{n}\mathrm{LO}$+$\mathrm{N}^{n}\mathrm{LL}$ predictions. The current picture of parton showering could be greatly improved. Quark-gluon discrimination studies would greatly benefit from large pure quark/gluon samples in the clean enviornment of a lepton collider, yielding much better discriminating power than in pp collisions. A $<1\%$ control of colour reconnection could be obtained by exploiting the large number of $\mathrm{e^{+}e^{-}}\rightarrow \mathrm{WW}$ events at the FCC-ee. Finally, high-precision hadronization studies could be carried out, with a focus on colour string dynamics.

%%%%%%%%%%%%%%%%%%%%%%%%%%%%%%%%%%%%%%%%%%%%%%%%%%%%%%%%%
\begin{center} \textbf{Acknowledgments} \end{center}
I would like to thank David d'Enterria for the input material for this talk, as well as many helpful discussions. I would also like to thank my supervisor Freya Blekman for helpful discussions and feedback during preparation.

%
%%%%%%%%%%%%%%%%%%%%%%%%%%%%%%%%%%%%%%%%%%%%%%%%%%%%%%%%%

\newpage
\bibliographystyle{JHEP}
\bibliography{main}

%%%%%%%%%%%%%%%%%%%%%%%%%%%%%%%%%%%%%%%%%%%%%%%%%%%%%%%%%%%%%%%%%%%%%%%%%%%

\end{document}